\documentclass{emulateapj}

\newcommand{\varcsec}{^{\prime\prime}}
\renewcommand{\arcsec}{.\hspace{-0.9mm}^{\prime\prime}\hspace{-0.2mm}}

\slugcomment{To appear in Astrophysical Journal Letters}

\shorttitle{Quiet-Sun intensity contrasts in the near ultraviolet}
\shortauthors{Hirzberger et al.}

\begin{document}


\title{Quiet-Sun intensity contrasts in the near ultraviolet as
  measured from Sunrise}


\author{\textsc{
J.~Hirzberger$^1$, 
A.~Feller$^1$,
T.L.~Riethm\" uller$^1$,
M.~Sch\" ussler$^1$,
J.M.~Borrero$^{2,1}$,
N.~Afram$^3$,
Y.C.~Unruh$^3$,
S.V.~Berdyugina$^2$,
A.~Gandorfer$^1$,
S.K.~Solanki$^{1,4}$,
P.~Barthol$^1$,
J.A.~Bonet$^5$,
V.~Mart\'\i nez Pillet$^5$,
T.~Berkefeld$^2$,
M.~Kn\" olker$^6$,
W.~Schmidt$^2$,
A.M.~Title$^7$}}
\affil{
$^1$Max-Planck-Institut f\" ur Sonnensystemforschung, D-37434
    Katlenburg-Lindau, Germany \\
$^2$Kiepenheuer-Institut f\" ur Sonnenphysik, D-79104 Freiburg,
              Germany \\
$^3$Astrophysics Group, Blackett Laboratory, Imperial College, 
              London, SW7 2AZ, UK \\
$^4$School of Space Research, Kyung Hee University, Yongin,
              Gyeonggi 446-71, Korea\\
$^5$Instituto de Astrof\'\i sica de Canarias, E-38200 La
              Laguna, Spain \\
$^6$High Altitude Observatory, National Center for Atmospheric Research$^{\ast}$, 
              Boulder, Colorado 80307, USA \\
$^7$Lockheed Martin Solar and Astrophysics Laboratory, Palo Alto,California
              94305, USA}

\altaffiltext{$\ast$}{The National Center for Atmospheric Research is sponsored by the National 
                 Science Foundation}


\email{hirzberger@mps.mpg.de}

\begin{abstract}
We present high-resolution images of the Sun in the near ultraviolet
spectral range between 214\,nm and 397\,nm as obtained from the first
science flight of the 1-m {\sc Sunrise} balloon-borne solar
telescope. The quiet-Sun rms intensity contrasts found in this
wavelength range are among the highest values ever obtained for
quiet-Sun solar surface structures -- up to 32.8\,\% at a wavelength
of 214\,nm. We compare with theoretical intensity contrasts obtained
from numerical magneto-hydrodynamic simulations. For 388\,nm and
312\,nm the observations agree well with the numerical simulations
whereas at shorter wavelengths discrepancies between observed and
simulated contrasts remain.
\end{abstract}





\section{Introduction}

Quiet-Sun intensity fluctuations provide fundamental information on
the thermal structure of the convective overshoot region at the solar
surface. A large number of measurements of the quiet-Sun intensity
fluctuations have been made in the visible spectral range, starting in
the 1950s \citep{frenkiel55} and reaching to comparisons between
recent seeing-free observations from the Solar Optical Telescope (SOT)
on the Hinode satellite, which provide contrast values comparable to
those from state-of-the-art numerical simulations
\citep[e.g.,][]{danilovic08,mathew09,wedemeyer09}. Overviews of the
results of former measurements are given by \citet{beckers69} and
\citet{sanchez00}. Although many intensity contrast measurements are
based on data obtained after careful correction of atmospheric and/or
instrumental effects, it turns out that insufficient knowledge of
optical aberrations and straylight still are the major obstacles for
accurate contrast measurements.

In the near ultraviolet (NUV), the intensity fluctuations of the solar
surface region have been hitherto largely unknown, owing to the lack
of reliable imaging data. The strong dependence of the Planck function
in the NUV to the temperature makes the intensity highly sensitive to
temperature fluctuations. Therefore, intensity contrasts also
represent important diagnostics to probe the validity of numerical
simulations of convection at granular scales.

From an observational point of view, it is well known that seeing
effects and straylight contributions increase towards shorter
wavelengths. Ground-based NUV observations are, additionally, hampered
by strong atmospheric absorption (mainly by stratospheric
ozone). Below the atmospheric cut-off at about 315\,nm
\citep[see][]{gandorfer05}, observations cannot be carried out from the
ground.

In the present study we show for the first time high-resolution
imaging data and intensity fluctuations in the NUV region down to
214\,nm. These data were obtained during the first science flight of
the balloon-borne 1-m {\sc Sunrise} observatory
\citep[see][]{barthol10,martinez10,berkefeld10,solanki10}. We present
disk-center quiet-Sun NUV intensity contrasts obtained with the {\sc
  Sunrise} Filter Imager \citep[SuFI, see][]{gandorfer10} onboard {\sc
  Sunrise}.

%
%

\section{Observations}

The first science flight of the balloon-borne {\sc Sunrise} telescope was
carried out between June~8 and June~13, 2009, from Kiruna, Sweden, to Somerset
Island, Canada. The average cruise altitude was about 36\,km, i.e., at the
upper boundary of the stratospheric ozone layer, so that high resolution
images in the NUV could be obtained. The spectral regions observed by SuFI are
214\,nm, 300\,nm, 312\,nm (close to the bandhead of the OH molecule), 388\,nm
(bandhead of the CN molecule), and 396.8\,nm (Ca\,II\,$H$ line core). The
widths (FWHM) of the corresponding SuFI filters are 10\,nm, 5\,nm, 1.2\,nm,
0.8\,nm, and 0.18\,nm, respectively. The passbands within CN and OH bandheads
were selected because of their high temperature sensitivity, especially UV OH
lines which are favorable for imaging of both the quiet photosphere and
sunspots \citep{berdyugina03}. The filters are mounted in a filter wheel that
allows for sequential observations of the five wavelength channels.

SuFI is equipped with a 2k\,$\times$\,2k CCD detector (PixelVision
BioXight) at an effective focal length of 121\,m. The camera pixel
pitch is 12\,$\mu$m corresponding to $0\arcsec 0207$ (at 300\,nm) on
the sky. For post-facto correction of low-order optical aberrations by
means of phase-diversity wavefront sensing (PD), SuFI was equipped
with a device providing a focussed and a defocussed image, each on
one half of the detector. The effective field of view (FOV) is thus
limited to about $15\varcsec\times 40\varcsec$.

The exposure times for the SuFI observations were strongly
wavelength-dependent and ranged from 75\,ms in the CN channel up to
30\,s in the 214\,nm channel. In this latter channel, the intensity
level was high enough for detection only around maximum solar
elevation and maximum balloon altitude. Therefore, observations in the
214\,nm channel were only carried out around local noon.

\begin{figure*}

\vspace*{8mm}

\includegraphics[height=18cm,angle=90]{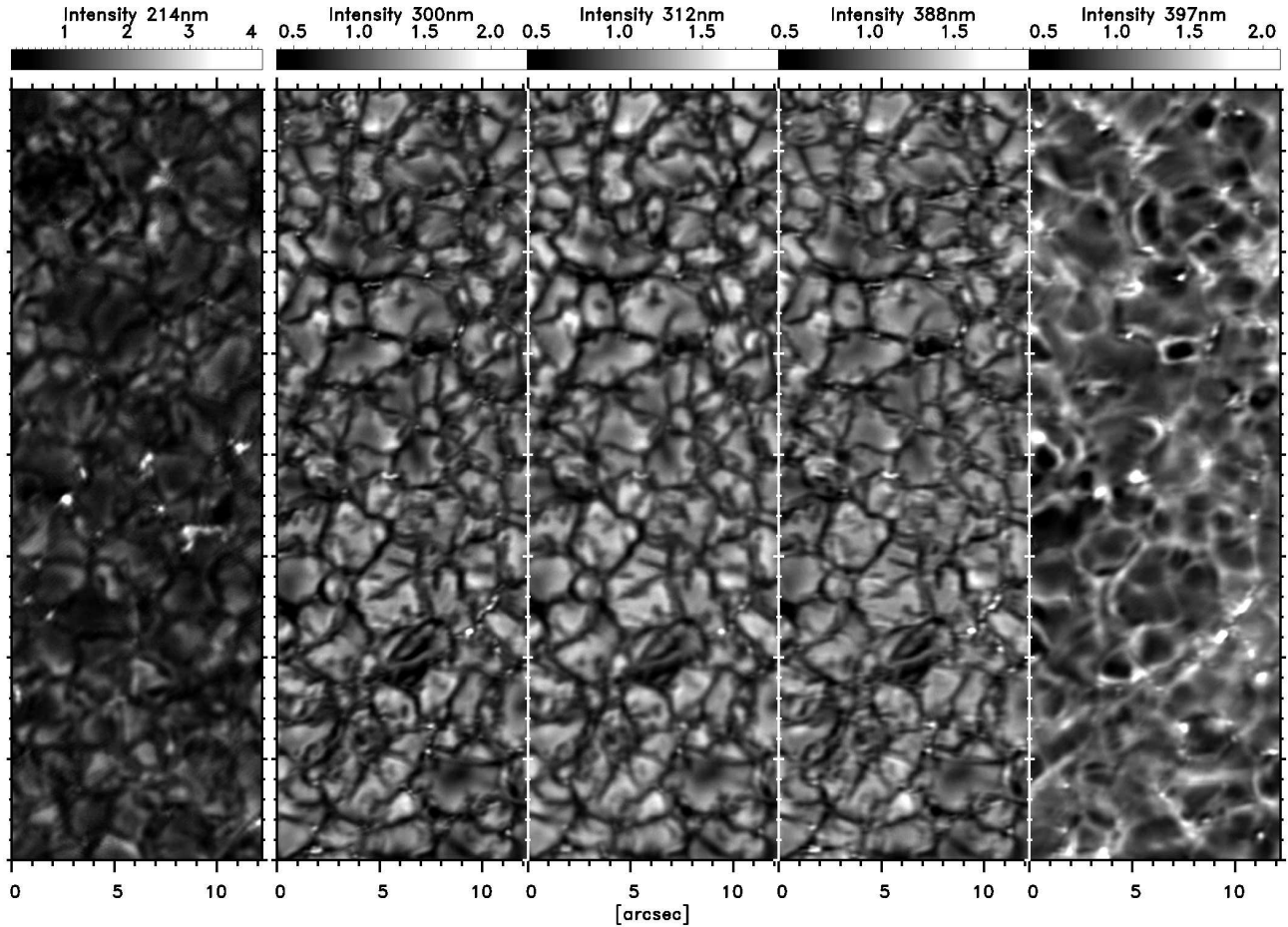}
\caption{\label{fig1} Examples of PD-reconstructed (level 2) images
  obtained in the 5 {\sc Sunrise}/SuFI wavelength channels. The
  displayed image in the leftmost panel (214\,nm) was obtained on
  June~9, 2009, 14:59:24\,UT, the other images on June~11,
  20:07:41\,UT (300\,nm), 20:07:42\,UT (312\,nm), 20:07:38\,UT
  (388\,nm), and 20:07:39\,UT (397\,nm). All images show fields of quiet
  granulation near disk center. The intensities have been normalized
  to the mean intensity $\bar I$ over the FOV and the scales have been
  limited to $\bar I\pm 10\sigma$ (214\,nm) and $\bar I\pm 5\sigma$
  (all other channels), respectively, where $\sigma$ denotes the
  standard deviation of the corresponding intensity.  }
\end{figure*}

Here we show data obtained on June~11, 2009, between 20:00\,UT and
21:00\,UT (300\,nm, 312\,nm, 388\,nm, and 397\,nm) and on June~9,
2009, between 14:00\,UT and 15:00\,UT (all SuFI channels).  Examples
of single images obtained in the 5 SuFI wavelength channels are shown
in Fig.~\ref{fig1}.

%
%

\section{Data analysis}\label{data}

The raw (``level~0'') SuFI images were corrected for dark and flat
fields. Dedicated flat-field images were obtained by moving the
telescope during exposure.  For flat field correction of the science
images accumulations of all flat-field and science images obtained in
an interval of $\pm 2$\,hours around the observation time were
used. In addition, clusters of bad detector pixels were eliminated by
applying a local median filter. Residual inhomogeneities originating
from scratches at the entrance window of the SuFI camera were removed
by local smoothing using low-pass filters.

The flat- and dark-field corrected SuFI images (henceforth labelled as
``level 1'' data) were reconstructed by means of a PD algorithm. We
used a PD code originally described by \citet{loefdahl94} and further
developed as described by \citet{bonet04} and \citet{vargas08}. In
order to optimize the reconstruction of the SuFI data, several
additional features were included, which will be described together
with a performance study of the new PD code in a forthcoming
publication \citep[][in prep.]{hirzberger10}.

In contrast to ground-based observations, the contribution of
atmospheric influences to the total wavefront aberrations at balloon
altitudes is small compared to instrumental effects. Therefore, we
expect isoplanatic conditions, i.e. constant wavefront aberration
across the SuFI FOV. Temporal variations of the wavefront deformations
are assumed to arise only from changes of the instrument's
temperatures and from slowly varying mechanical deformations, which
both are mainly correlated with the solar elevation. These assumptions
allow a homogeneous reconstruction of the SuFI images. We thus
included a procedure, which permits averaging the fitted Zernike
coefficients across the entire FOV as well as over all images within a
certain time span (usually one hour). Subsequently, all images taken
during this time span were reconstructed with constant and highly
reliable wavefront errors. PD reconstruction of individual images is,
henceforth, denoted as ``level~2'' reduction, while images
reconstructed with averaged wavefront errors are labelled as
``level~3''.

%
%

\section{Results}

\subsection{Disk-center intensity contrasts}

\begin{figure*}
\includegraphics[width=16cm]{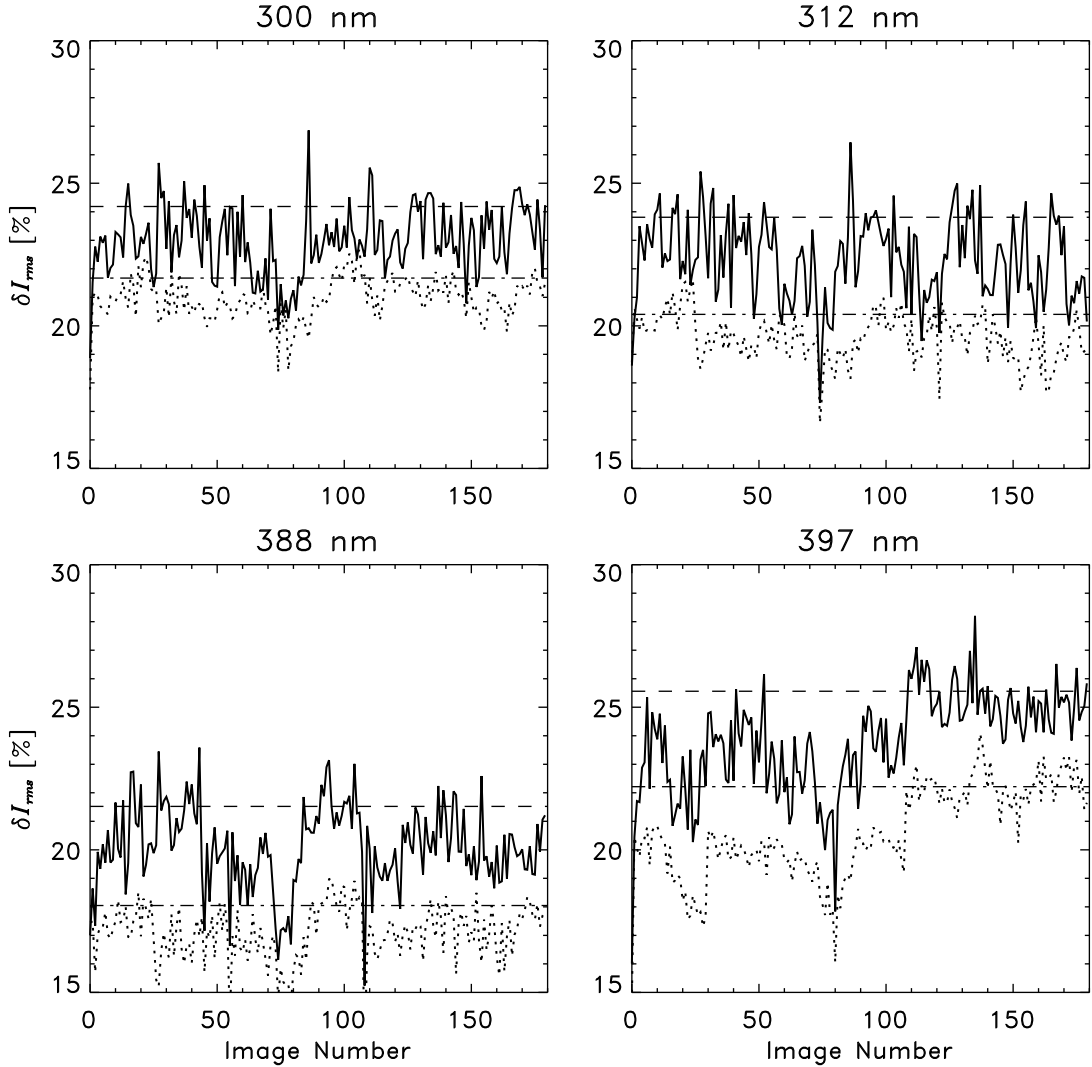}
\caption{\label{fig2} rms intensity contrasts of 180 images in each
  of four wavelength channels, obtained on June~11, 2009, between
  20:00 and 21:00~UT. Solid lines correspond to level-2 data and
  dotted lines refer to level-3 data. The dashed and dash-dotted
  horizontal lines lie one standard deviation above the mean contrast
  of the level-2 and level-3 images, respectively. In the 397\,nm
  band, the contrast is slightly higher in the second half of the
  image series than in the first half since the FOV had moved by a few
  arcseconds on the solar disk.}
\end{figure*}

The quiet-Sun disk-center rms intensity contrasts, $\delta I_{rms}$,
derived from the data obtained on June~11, 2009 are plotted in
Fig.~\ref{fig2}. In each spectral channel, for 180 consecutive images
(approximately 24\,min) the image stabilization system \citep[{\sc
    Sunrise} Correlating Wavefront Sensor, CWS; see][]{berkefeld10}
was locking in closed loop. The obtained intensity contrasts show
considerable temporal fluctuations which are partially caused by the
evolution of the solar structures and p-modes but also by a variation
of the image quality. The image quality is affected, first, by the
limited overall stability given by the gondola pointing and the image
stabilization in presence of wind gusts and vibrations
\citep[][]{berkefeld10}, and, secondly, by residual artifacts, such as
scratches and bad pixels in the level-1 images (see Sect.~\ref{data}),
which are amplified by the PD reconstruction. After visual inspection,
the influence of the latter is considered to be negligible in the data
presented here.

In addition to the temporal fluctuations, a rather constant offset of
approximately $2-3$\,\% (depending on wavelength) of the intensity
contrasts from level-2 to level-3 data is evident. This offset is
expected since for the level-2 data always the local best fits to the
wavefront aberrations are used for reconstruction (which also may cause
local over-reconstructions), whereas the averaged wavefronts used for
reconstructing the level-3 data may locally underestimate the
wavefront errors (e.g. due to a small amount of anisoplanatism).

Irrespective of the effect of the image processing methods, the
photospheric intensity contrasts shown in Fig.~\ref{fig2} are among
the highest quiet-Sun contrasts ever measured. The contrast variations
within the time series are to a lesser extent related to granular
evolution than to the varying image quality due to residual pointing
errors. This sometimes leads to artifacts due to local
over-reconstruction and thus to unrealistically high contrast
values. In the majority of images this effect, however, tends to
result in reduced intensity contrasts (by smearing out solar surface
structures). In order to allow for this systematic reduction of
contrasts and to clip outliers, we consider the
``mean-plus-one-sigma'' values, $\left<\delta I_{rms}\right> +
\sigma$, as reliable {\it maximum} contrast measurements. Here,
$\left<\delta I_{rms}\right>$ denotes the temporally averaged rms
contrast and $\sigma$ denotes the corresponding standard
deviation. The mean-plus-one-sigma values are overplotted in
Fig.~\ref{fig2} and given in Tab.~\ref{tab1}.

\begin{figure*}
\includegraphics[width=16cm]{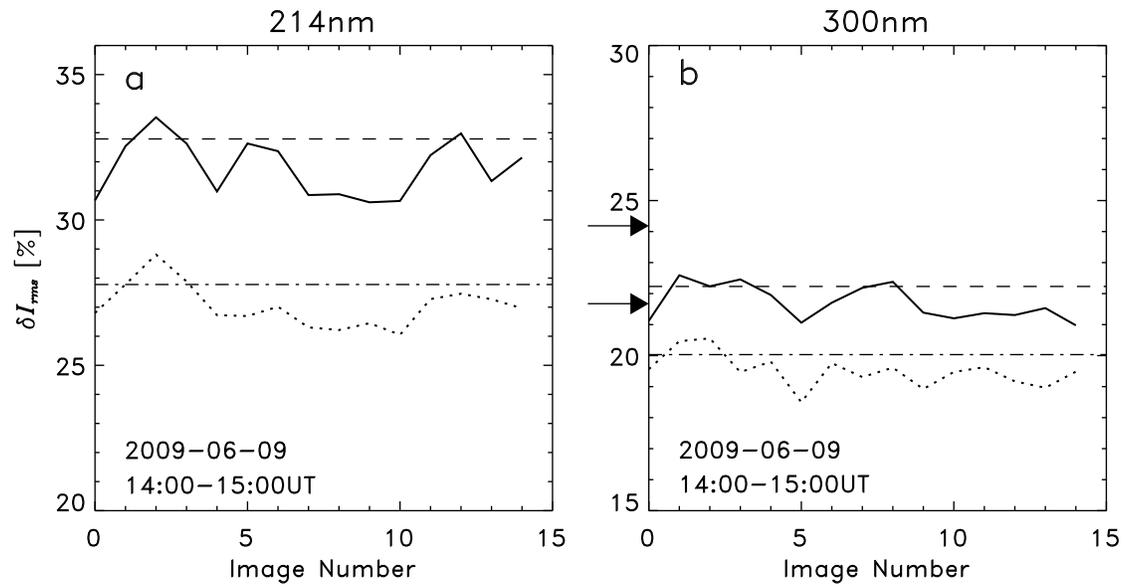}

\vspace*{5mm}

\caption{\label{fig3} rms intensity contrasts of 15 images taken in
  the 214\,nm (panel a) and 300\,nm (panel b) wavelength channels,
  obtained on June~9, 2009, between 14:00 and 15:00 UT. Solid lines
  correspond to level-2 data and dotted lines refer to level-3
  data. The dashed and dash-dotted horizontal lines show the
  mean-plus-one-sigma values of level-2 and level-3 images,
  respectively. For comparison, the mean-plus-one-sigma values
  obtained from the June~11 300\,nm data are indicated by arrows.  }
\end{figure*}

Figure~\ref{fig3} shows the intensity contrasts of 15 images from the
214\,nm and 300\,nm channels of the June~9 data. As a result of the
long exposure time at 214\,nm, the number of reliable images obtained
during the observing period is limited to that small number. The
mean-plus-one-sigma rms contrasts of the June~11 data are marked by
arrows. Due to higher residual pointing errors, which caused some
smearing of the images, the June~9 data are of slightly worse quality
than the June~11 data. This particularly affects the 214\,nm data.
Nevertheless, the obtained intensity contrasts in the 214\,nm channel
are the highest photospheric quiet-Sun contrasts ever measured.

\subsection{Intensity contrasts from simulations}

In order to compare with the measured intensity contrasts, we have
calculated NUV intensity maps from numerical magneto-hydrodynamic
(MHD) simulations of the quiet solar photosphere \citep{voegler05}. We
used snapshots from a simulation run with a mean vertical magnetic
field strength of $\left< B_z\right>$ = 50\,G and a horizontal cell
size of 20.8\,km. The intensity maps were calculated by using the
spectral synthesis code ATLAS9 \citep{kurucz93} which specifies the
opacity distribution functions \citep[ODFs,][]{strom66} in the SuFI
300\,nm, 312\,nm, and 388\,nm channels. With the current
implementation for computing ODFs, we are unable to calculate reliable
intensities below a wavelength of 220\,nm.  In order to estimate the
simulated contrast at 214\,nm, we tentatively applied the ODF for
220\,nm to the MHD snapshots. In addition, in the SuFI 388\,nm channel
a full spectral synthesis with the SPINOR code
\citep{solanki87,frutiger00,berdyugina03}, including all available
atomic and molecular line parameters within the SuFI 388\,nm spectral
bandpass have been performed.

\begin{figure}
\includegraphics[width=8cm]{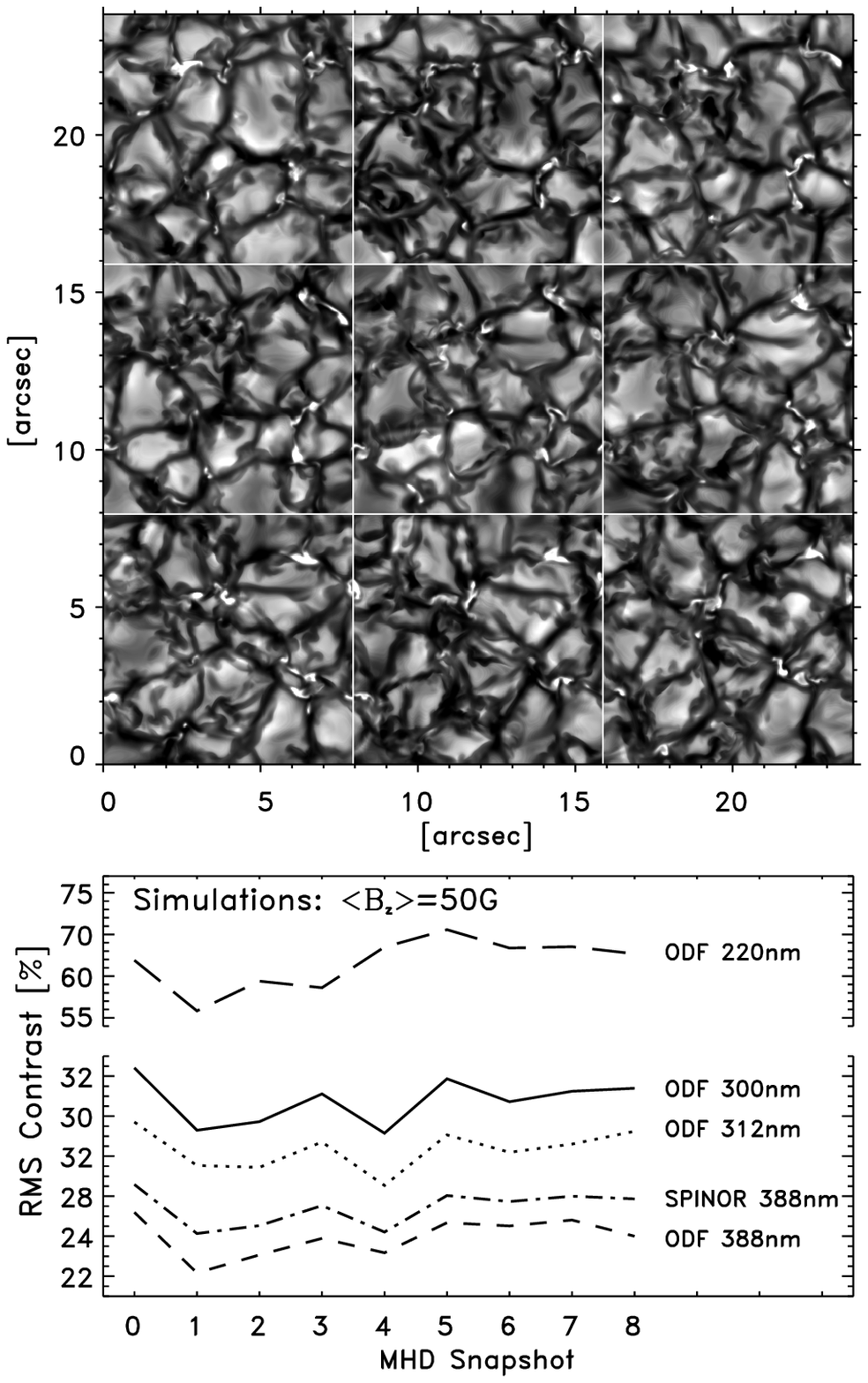}
\caption{\label{fig4} Upper panel: Nine brightness images at 300\,nm
  from a MHD simulation with a mean vertical field strength of $\left<
  B_z\right> = 50$\,G; lower panel: corresponding $\delta I_{rms}$
  calculated for four of the five SuFI spectral channels, using two
  different methods for 388\,nm (ODF-based and full spectral
  synthesis). Since ODFs do not allow reliable intensity calculations
  at 214\,nm, the computations have been carried out at the nearest
  wavelength at which they are better established, 220\,nm.}
\end{figure}

Nine synthetic intensity images at 300\,nm and the corresponding
$\delta I_{rms}$ for all SuFI channels are shown in
Fig.~\ref{fig4}. Owing to the small area (6\,Mm$\times 6$\,Mm) of the
MHD box, the simulated contrasts show a temporal variation with a
standard deviation of about 1\,\% (except for the 220\,nm spectral
region where the value is about 3\,\%). The contrasts obtained with
SPINOR in the 388\,nm channel are consistently about 1.3\,\% higher
than those obtained by using ODFs.  This small difference, comparable
to the temporal variation of $\delta I_{rms}$, indicates that the
ODF-based spectral synthesis in the other spectral channels may not be
completely unrealistic, although NLTE effects have not been taken into
account. The mean contrasts obtained from the MHD simulations are also
given in Tab.~\ref{tab1}.

%
%

\section{Discussion and conclusions}

The NUV intensity contrasts obtained from {\sc Sunrise}/SuFI data are
systematically lower than the values from MHD simulations by several
percent (but nearly a factor of two at 214\,nm). Since the spatial
resolution of both datasets is similar and because the instrumental
aberrations have been removed by means of PD reconstruction, three
possible sources for the discrepancy remain: (i) inaccurate physics in
the simulations, (ii) omission of NLTE effects in the intensity
calculations, (iii) scattered light. Before we can estimate the
importance of NLTE effects or the need to improve the simulations, we
need to judge the influence of scattered light on the measured
contrasts.

The amount of scattered light in the SuFI data has not yet been fully
assessed. A first estimate for the 300\,nm, 312\,nm and 388\,nm SuFI
channels was derived from limb observations. The straylight components
of the modulation transfer function obtained from limb profiles at the
different wavelengths were applied to the numerical simulations to
estimate the effect of straylight on the rms intensity contrasts (see
Tab.~\ref{tab1}; details will be presented by \citet[][in
  prep.]{feller10}). For 214\,nm and 397\,nm we were not able, yet, to
obtain a straylight estimate due to the lack of limb observations
(214\,nm) and due to the ubiquitous presence of spicules (397\,nm). At
312\,nm and 388\,nm, the synthetic images after stray light
contamination display contrast values that lie between the level-2 and
the level-3 data. At these wavelengths, the simulations thus give
results consistent with the measurements. At 300\,nm the simulations
give a value of 25.5\,\% that is still 1.3\,\% higher than the 24.2\,\%
obtained from the level-2 data.

Estimates of scattered light contributions in (also seeing free) data
obtained with Hinode/SOT were carried out by \citet{mathew09} and
\citet{wedemeyer09}. They found a significant increase of the
intensity contrasts in the visible Hinode channels, after deconvolving
with point-spread functions for scattered light. For the 388\,nm
spectral region \citet{mathew09} found a quiet-Sun intensity contrast
of 21.8\,\% after removing scattered light, which is very similar to
the mean-plus-one-sigma value of our level-2 data from June~11, 2009
(cf. Tab.~\ref{tab1}). Therefore, we conclude that after applying
straylight correction to the SuFI-388\,nm data, the obtained contrasts
will exceed those obtained by \citet{mathew09} from Hinode data and
will be comparable to synthetic contrasts achieved from MHD data.

The consistency between the measured and the numerically simulated
contrasts indicates that the temperature fluctuations in the lower
photosphere are correctly described by the hydrodynamical simulations,
which predict values of $\delta T_{rms}$ between 2\,\% and 5\,\% on
surfaces of constant optical depth (at 500\,nm) between $\tau_{500}=1$
and $\tau_{500}=0.01$. Taken together with the reproduction of shifts
and asymmetries of spectral lines \citep[e.g.][]{nordlund09} and
inverse granulation \citep{cheung07} by simulation results, our
contrast data thus provide evidence for an extension of the convective
overshoot to a height of roughly 300\,km above the average level of
optical depth unity, at which height the rms fluctuations of the
vertical velocity reach a local minimum after a steep decrease (by
about a factor of 4) from their maximum slightly below the visible
surface (optical depth unity).

Direct translation of rms contrasts into e.g. temperature
stratification of the solar photosphere can be carried out only by
means of simplified models estimating the brightness temperature of
atmospheric features \citep[e.g.][]{solanki98,sobotka00}. In the NUV
the rms intensity contrast of broad band images as obtained with
{\sc Sunrise}/SuFI is, however, only partly a function of height. It
depends just as much on the temperature sensitivity of the Planck
function and of the lines in the SuFI passbands, the atomic species,
ionization stages and in the case of molecular transitions, the
dissociation energy of the molecules.

In summary, we conclude that the data obtained from the first science
flight of {\sc Sunrise}/SuFI are of resounding quality. They show the
highest quiet-Sun photospheric intensity contrasts ever measured, even
without correcting for scattered light.  


%
%

\begin{table*}
\caption{\label{tab1} Mean-plus-one-sigma values of $\delta I_{rms}$
  (see Figs.~\ref{fig2} and \ref{fig3}) and mean $\delta I_{rms}$
  resulting from MHD simulation data (in percent). In the rightmost
  two columns, mean $\delta I_{rms}$ of MHD data convolved with
  preliminary estimated levels of scattered light are given.}

\begin{tabular}{rrrrrrrrr}
\hline
  $\lambda$      & \multicolumn{2}{c}{June 9} & \multicolumn{2}{c}{June 11} &
          \multicolumn{2}{c}{MHD} &
          \multicolumn{2}{c}{MHD straylight}\\
  $\left[\mbox{nm}\right]$      & level 2 & level 3 & level 2 & level 3 & ODF & SPINOR & ODF & SPINOR\\
\hline
214 &  32.79 ($31.80+0.98$) & 27.78 ($27.05+0.73$) &     - &     - & 61.27$^{\ast}$           &     - &     - &     - \\
300 &  22.23 ($21.63+0.60$) & 20.03 ($19.52+0.52$) & 24.19 ($22.98+1.21$) & 21.67 ($20.94+0.73$) &
30.76\phantom{$^{\ast}$} &     - & 25.50 &    \\
312 &  21.91 ($20.84+1.07$) & 19.45 ($18.79+0.66$) & 23.81 ($22.34+1.47$) & 20.40 ($19.58+0.82$) & 28.34\phantom{$^{\ast}$} &     - & 22.03 &     - \\
388 &  19.16 ($18.28+0.89$) & 17.27 ($16.73+0.55$) & 21.52 ($20.09+1.43$) & 18.05 ($17.11+0.94$) & 23.94\phantom{$^{\ast}$} & 25.27 & 18.60 & 19.93 \\
397 &  22.64 ($21.33+1.30$) & 20.11 ($19.19+0.92$) & 25.56 ($23.81+1.75$) & 22.22 ($20.60+1.61$) &    - \phantom{$^{\ast}$} &     - &     - &     - \\
\hline
\end{tabular}\\
$^{\ast}$Mean $\delta I_{rms}$ at 220\,nm.
\end{table*}

%
%

%
%

%
%

%
%

%
%
\acknowledgments The German contribution to {\sc Sunrise} is funded by
the Bundesministerium f\"{u}r Wirtschaft und Technologie through
Deutsches Zentrum f\"{u}r Luft- und Raumfahrt e.V. (DLR), grant number
50~OU~0401, and by the Innovationsfond of the President of the Max
Planck Society (MPG). The Spanish contribution has been funded by the
Spanish MICINN under projects ESP2006-13030-C06 and AYA2009-14105-C06
(including European FEDER funds). The HAO contribution was partly
funded through NASA grant number NNX08AH38G. This work has been
partially supported by the WCU grant (No. R31-10016) funded by the
Korean Ministry of Education, Science and Technology. NA and YCU
acknowledge the NERC SolCli consortium grant. SVB acknowledges the 
EURYI (European Young Investigator) Award provided by the ESF 
(see www.esf.org/euryi) and the SNF grant PE002-104552.

%
%

\end{document}